\documentclass[a4paper]{jpconf}
\usepackage{graphicx}
\usepackage{caption}

\begin{document}
\title{On identifying the neutron star that was born in the supernova that placed $^{60}$Fe onto the Earth}

\author{R Neuh\"auser, N Tetzlaff, T Eisenbeiss, M M Hohle}

\address{Astrophysikalisches Institut, University Jena,
Schillerg\"a\ss chen 2-3, D-07745 Jena, Germany}

\ead{rne@astro.uni-jena.de}

\begin{abstract}
Recently, $^{60}$Fe was found in the Earth crust formed in a 
nearby recent supernova (SN). If the distance to the SN
and mass of the progenitor of that SN was known,
then one could constrain SN models.
Knowing the positions, proper motions, and distances of
dozens of young nearby neutron stars,
we can determine their past flight paths and possible kinematic origin. 
Once the birth place of a neutron star in a SN is found,
we would have determined the distance of the SN
and the mass of the SN progenitor star. 
\end{abstract}

\section{Introduction: Link between 60Fe as SN debris on Earth and Neutron Stars}

$^{60}$Fe was found in the Earth crust,
which arrived on Earth $\sim 2$ Myr ago (Knie et al. 2004).
Production of $^{60}$Fe in red giants cannot explain the observed peak,
so that its origin can only be a nearby (40-100 pc) supernova (SN).
If one could estimate the mass ejected by the SN
and the mass of the progenitor star, one could constrain SN models.
To estimate the amount of $^{60}$Fe ejected, one needs to know the
uptake factor (fraction of material deposited in the crust among the whole 
material arrived on Earth), the SN distance, the half-life of $^{60}$Fe,
and the time since the SN
(i.e. the sum of the time mentioned above, $\sim 2$ Myr, since the deposition
in the crust, plus the travel time from the SN to the Earth's crust through interstellar 
space and Earth atmosphere).
The uptake factor is probably few per cent, but hardly known with high precision;
it is possible to constrain the factor, if other material would also be found
in the Earth's crust from the same SN event.
The travel time from the SN to the Earth crust is most likely negligible compared to
the $\sim 2$ Myr, since the SN must have been nearby, roughly 40 pc
(within a factor of 2): If the SN would have been more distant
than $\sim 100$, the SN ejecta would not have been able to penetrate into the
Solar System given the Solar wind ram pressure; if the SN would have been
closer than $\sim 10$ pc, a large cosmic ray flux on Earth would have caused
major climatic changes and a large mass extinction of life.
Hence, the distance towards the SN remains uncertain within the limits given,
so that one cannot derive the amount of $^{60}$Fe ejected from the amount found on Earth.
Also, the mass of the SN progenitor star remains unknown.

If we can identify the neutron star (NS) born in that SN
and, by tracing back the motion of that NS, also the location of that SN event, 
we would know the distance towards the SN, i.e. could solve the degeneracy.
We may even be able to determine the mass of the progenitor star
from the difference between the age of the parent association (if the 
NS was born in an association) and NS kinematic age (flight time),
if all stars were formed at once in that association, as it is assumed
by star formation theory and observed for most clusters.
Such results would be very important and valuable for SN theory,
ejecta models, nucleosynthesis, etc.

\section{Tracing back neutron stars to their birth place in a supernova}

NSs are born when massive stars ($\sim 8$ to 30 M$_{\odot}$) explode as SNe
after their short life-times of a few Myr.
Hence, many of them should still be embedded in their star forming region,
stellar association, or cluster, when they explode as SNe. 
In such SNe, NSs are formed, which receive a kick and fly away with large space motion.
Given the measured 2D proper motions of NSs, one can construct the likely 3D (and 1D radial
velocity) distributions (Arzoumanian et al. 2002, Hobbs et al. 2005).
NSs also cool down fast after the SN, but are still hot enough for detection
as faint optical or bright X-ray sources for up to 5 Myrs (see cooling curves 
in Gusakov et al. 2005, Popov et al. 2006).
Young NSs are detectable (and have been detected) as optical and/or X-ray sources within
1 kpc (e.g. review in Haberl 2007).
If the SN takes takes place in a binary system, then the companion star gets ejected
and may then be observable as a run-away star (Blaauw 1961); the exact direction and
velocity depends on the former binary orbit and the location of the NS and the companion
during the SN.
In principle, it is possible to trace back the motion of a NS to find out,
whether it may have flown through a star forming region or young stellar
association. If that is the case, and if the present mass function of that association 
is consistent with at least one very massive star, i.e. with at least one past SN, 
then the NS may have been born in that association. To get additional evidence for
such a scenario, and also in case that a particular NS could have flown through more
than one association, it would be best, if one could also identify a run-away star
as former companion of the SN progenitor, also to have been at the same time at the
same location as the NS and the association. Then, the place of origin and
the age of the NS would be known, as well as the distance towards the SN.
It is indeed possible to trace back both NSs and run-away stars to find close
possible encounters in space and time, i.e. evidence for a SN event in a binary,
e.g. Hoogerwerf et al. (2001), Tetzlaff et al. (2011b). Even without the identification of a former companion, the 
parent association of a NS can be found (Tetzlaff et al. 2009, 2010, 2011b).
Hence, it might be possible, but not guaranteed, to find the NS that 
was formed in the SN that also placed $^{60}$Fe onto Earth.

\section{Tracing back neutron star RXJ0720 and a run-away star as former companion}

To identify potential parent SN sites of a NS,
we calculate the trajectories backwards into the past
for the NS, the centre of all nearby young associations,
and all known young nearby run-away stars.
The sample of known young nearby NSs can be obtained from
the ATNF database\footnote{http://www.atnf.csiro.au/research/pulsar/psrcat/} (Manchester et al. 2005).
A list of all known young nearby associations has been put
together in Tetzlaff et al. (2010, 2011b).
For a catalog of nearby run-away stars, see Tetzlaff et al. (2011a).
For details on the procedure, see Tetzlaff et al. (2010, 2011b).
For run-away stars and associations, distances and 3D space motions
are known. For NSs, distances and 2D proper motions on sky are known,
but the radial velocity is unknown for most NSs.
Hence, we perform Monte-Carlo simulations
by drawing the radial velocity from the most likely distribution.

Recently, Eisenbeiss (2011) re-determined the parallaxe of the NS
RXJ0720 using archival Hubble Space Telescope data yielding $\pi = 3.6 \pm 1.6$ mas
(Eisenbeiss 2011), i.e. $\sim 278$ pc (192-500 pc); this is consistent
with an earlier determination by Kaplan et al. (2007),
(270-530 pc) within large error bars. We use the parallaxe and proper motion from 
Eisenbeiss (2011) and vary the unknown radial velocity $v_r$ in a Monte-Carlo simulation
using (a) the known distribution of NS velocities and, after we found that $v_r$ for 
RXJ0720 is probably relatively small, (b) a flat
distribution from $-300$ to $+300$ km/s).

There are nine possible parent associations, through which RXJ0720 could
have flown, i.e. in which it could have been born (TWA, Tuc-Hor, $\beta$ Pic-Cap,
HD 141569 group, AB Dor group, Col 140, Tr-10, CarA, Argus).
Then, there are also three possible former run-away companion stars
(i.e. former companions of the SN progenitor star),
which could have been at the same place at the same time as RXJ0720, for a certain NS radial velocity,
namely HIP 43158, 57269, and 76304 (run-away stars from Tetzlaff et al. 2011a).
If HIP 57269 or 76304 would have been the former companion, the SN would have
taken place at $\sim 27$ or $\sim 52$ pc, respectively, but the current measured distance of 
RXJ0720 would be inconsistent with the predictions in such cases.
For HIP 43158, however, the current measured distance of RXJ0720 is fully consistent
with the prediction in case that it would have been the former companion of the
RXJ07020 SN progenitor: The SN would have been $\sim 1$ Myr ago in Tr-10
at the distance of $\sim 370$ pc (during the SN), Fig. 1. Then, RXJ0720 should now
have a radial velocity of $\sim -100$ km/s. The difference between the kinematic
age of the NS ($\sim 1$ Myr) and the Tr-10 association age (15--35 Myr, Tetzlaff et al. 2010 
and references therein) then yield
the life-time and mass of the progenitor star, namely $\sim 13$ to 15 M$_{\odot}$.
Hence, RXJ0720 is not the NS born in the SN that placed the detected Fe onto Earth.
See Tetzlaff et al. (2011b) for details.

\begin{figure}
\centering
\includegraphics[height=3cm]{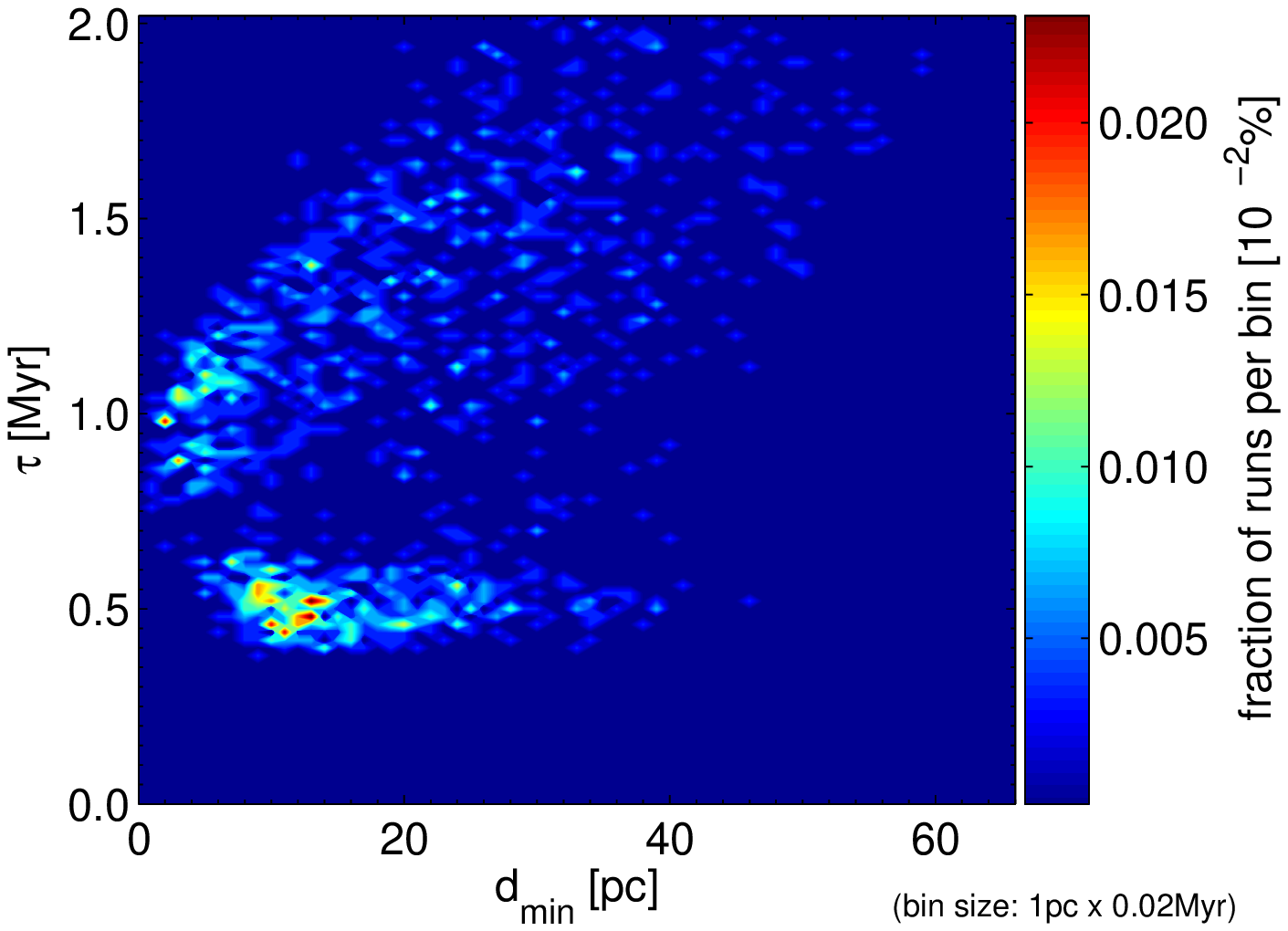}\nolinebreak
\includegraphics[height=3cm]{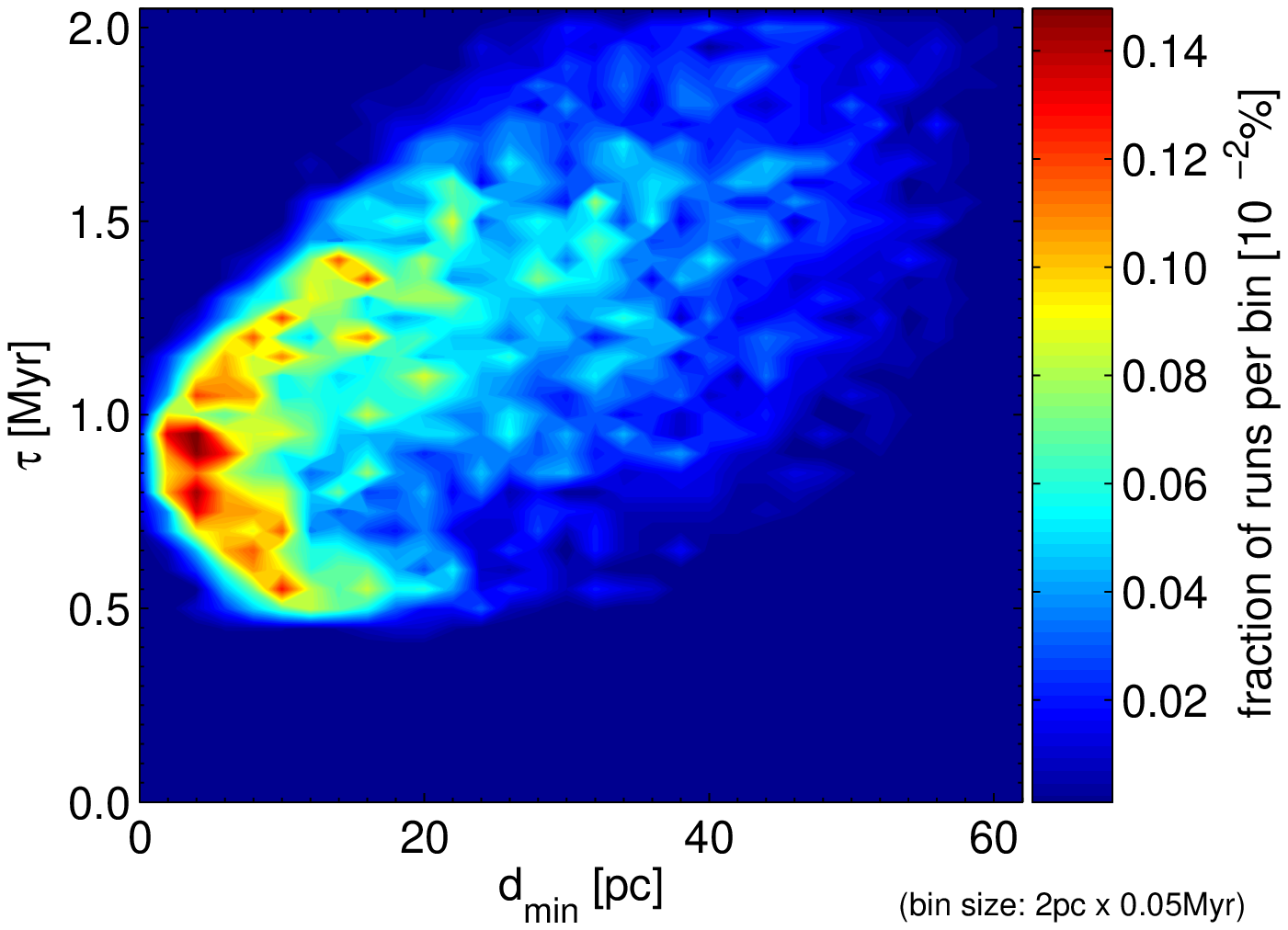}\nolinebreak\hspace{2em}
\fbox{\includegraphics[height=3cm]{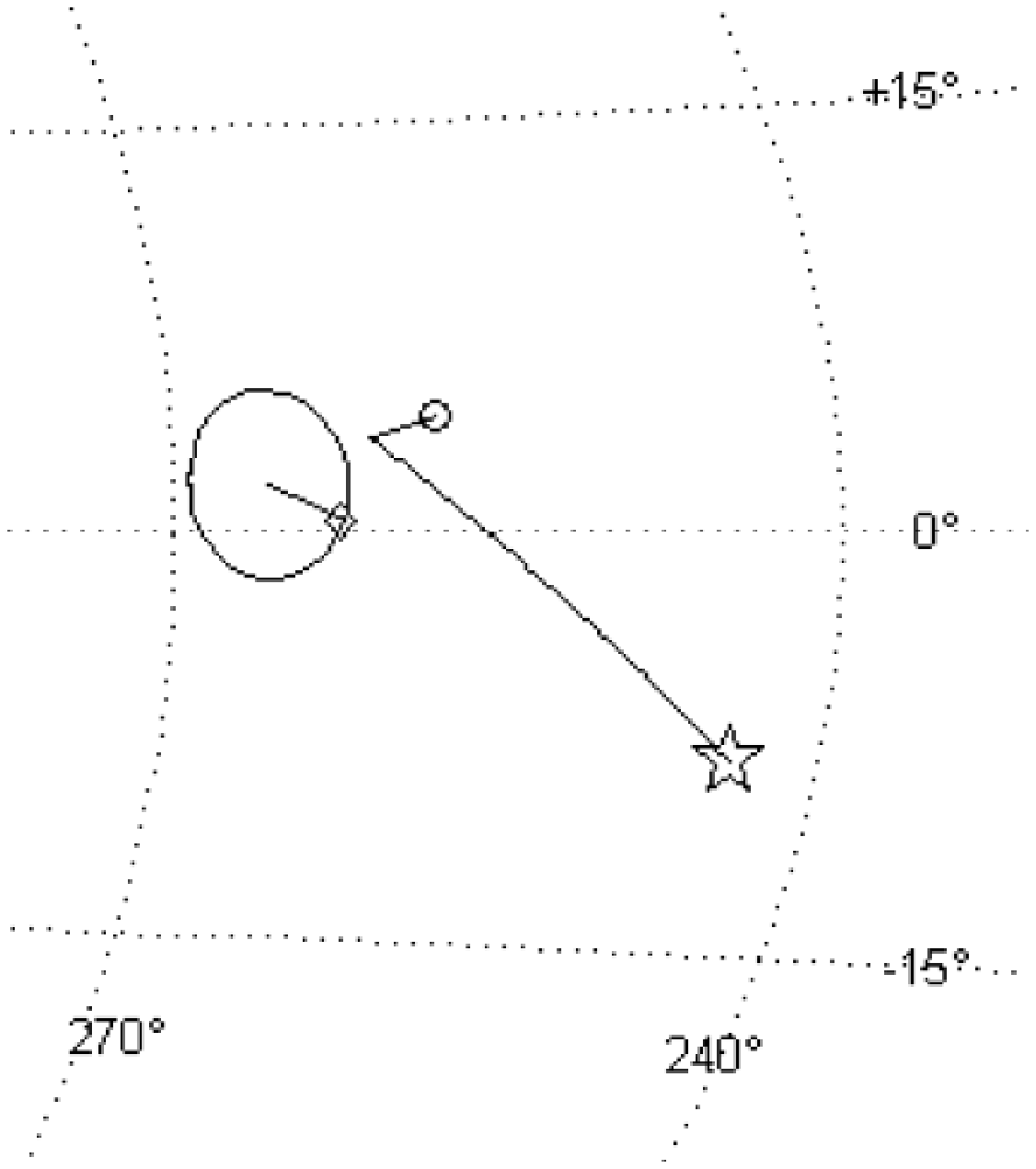}}
\caption{Contour plots for age (t in Myr since SN, left) and separation 
(d$_{\rm min}$ in pc, middle) distributions for the two cases, 
where NS RXJ0720 and HIP 43158 as run-away star (with radial velocity from
Hobbs et al. 2005, left, and for a flat distribution from $-300$ to $+300$ km/s, middle) 
could have been at the same place (in the Tr-10 association) at the same time in the past,
namely $\sim 0.7$ to $\sim 1$ Myr ago (Tetzlaff et al. 2011b). 
Right: Past trajectories ($\sim 1$ Myr, full lines) for RXJ0720 (star), 
the run-away star HIP 43158 (small circle),
and Tr 10 (diamond, radius 23 pc as large circle) projected on a
Galactic coordinate system (for a particular set of input parameters). 
Present positions are marked with the symbols.}
\end{figure}

\subsection{Acknowledgments}
We would like to thank DFG in SFB TR 7 and Carl-Zeiss-Stiftung for financial support.

\end{document}